\documentclass[aps,pre,reprint,twocolumn,superscriptaddress,showpacs]{revtex4-1}
\usepackage{amssymb,amsmath}
\usepackage[table]{xcolor}
\usepackage{graphicx}
\usepackage{mathtools}
\usepackage[export]{adjustbox}
\usepackage{overpic}
\usepackage[colorlinks=true,
 linkcolor=black,
 urlcolor=blue,
 citecolor=blue]{hyperref}
\usepackage{nicefrac}
\usepackage{multirow}
\usepackage{lipsum} 
\usepackage[normalem]{ulem}
\usepackage{pbox}
\newcolumntype{C}[1]{>{\centering\let\newline\\\arraybackslash\hspace{0pt}}m{#1}}
\usepackage{verbatim}
\usepackage{enumitem}

\usepackage{framed,color}
\definecolor{shadecolor}{rgb}{0.85,0.80,0.80}

\definecolor{myorange}{RGB}{253, 184, 99}
\definecolor{mypurple}{RGB}{178, 171, 210}

\newcommand{\comments}[1]{}
\usepackage{multirow}

\usepackage{float}

\newcommand{\beq}{\begin{equation}}
\newcommand{\eeq}{\end{equation}}
\newcommand{\bal}{\begin{aligned}}
\newcommand{\eal}{\end{aligned}}

\newcommand{\be}{\begin{equation}}
\newcommand{\ee}{\end{equation}}
\newcommand{\bd}{\begin{displaymath}}
\newcommand{\ed}{\end{displaymath}}
\newcommand{\BE}{\begin{eqnarray}}
\newcommand{\EE}{\end{eqnarray}}

\newcommand{\id}{{\openone}}

\newcommand{\boldpsi}{{\mbox{\boldmath $\psi$}}}
\newcommand{\boldphi}{{\mbox{\boldmath $\varphi$}}}

\allowdisplaybreaks
\begin{document}
\title{Persistent individual bias in a voter model with quenched disorder }
\author{Joseph W. Baron}
\email{josephbaron@ifisc.uib-scic.es}
\affiliation{Instituto de F{\' i}sica Interdisciplinar y Sistemas Complejos IFISC (CSIC-UIB), 07122 Palma de Mallorca, Spain}

\begin{abstract}
Many theoretical studies of the voter model (or variations thereupon) involve order parameters that are population-averaged. While enlightening, such quantities may obscure important statistical features that are only apparent on the level of the individual. In this work, we ask which factors contribute to a single voter maintaining a long-term statistical bias for one opinion over the other in the face of social influence. To this end, a modified version of the network voter model is proposed, which also incorporates quenched disorder in the interaction strengths between individuals and the possibility of antagonistic relationships. We find that a sparse interaction network and heterogeneity in interaction strengths give rise to the possibility of arbitrarily long-lived individual biases, even when there is no population-averaged bias for one opinion over the other. This is demonstrated by calculating the eigenvalue spectrum of the weighted network Laplacian using the theory of sparse random matrices.
\end{abstract}

%\pacs{05.10.Gg, 02.50.Ey, 87.10.Mn, 87.18.Cf}
%02.50.Ey Stochastic processes, 02.50.Ga Markov processes, 05.10.Gg	Stochastic analysis methods (Fokker-Planck, Langevin, etc.), 87.10.Mn Stochastic modeling, 87.18.Cf Genetic switches and networks

\maketitle

\section{Introduction}
The celebrated voter model \cite{clifford1973model, holley1975ergodic} offers a simple mathematical representation of opinion propagation. The parsimonious nature of the voter model provides a neutral canvas to which one can introduce more realistic features of social interactions. By introducing additional aspects to the original voter model, one can then observe their influence on idea propagation and on the possibility of consensus. Many such features have been studied including network architecture \cite{suchecki2005, vazquez2008analytical, sood2008voter}, the presence of zealots \cite{mobilia2007} or contrarians \cite{masuda2013, khalil2019noisy}, spontaneous change of opinion (or `noise') \cite{kirman, granovsky, carro2016noisy} and ageing \cite{peralta2020ordering}. The efficacy of the voter model in replicating real-world sociological data has also been investigated \cite{castellano2009statistical, fernandez2014}. 

Being a binary state stochastic system, the voter model has a strong analogy with Ising-type spin systems and is thus amenable to similar methods of statistical analysis. The universality class \cite{dornic2001critical}, the coarsening process \cite{suchecki2005}, statistically conserved quantities \cite{suchecki2004conservation, ben1996coarsening} and finite-size scaling are all well-understood \cite{krapivsky2010kinetic, liggett2012interacting}. 

Following the analogy with magnetic spin systems, a natural extension to the voter model is the introduction of quenched disorder. Quenched randomness is ubiquitous in models of spin glasses where it is employed to model the disordered couplings between spins \cite{mezard1987,edwards1975theory, sherrington}. In the context of the voter model, the effect of quenched disorder in the biases or imperturbabilities of individual voters (rather than the links between individuals) has been studied \cite{masuda2010, lafuerza, borile2013effect}. Selection rates weighted by the degrees of the nodes \cite{baronchelli, schneider2009generalized} have also been considered. Conversely, the effect of quenched disorder in link strength and the spin-glass transition have been studied in both the majority vote \cite{krawiecki2018majority, krawiecki2020ferromagnetic} and Glauber-Ising models \cite{crisanti1988dynamics, krawiecki2018ising}. 

In this work, we investigate the effect that quenched disorder, in the form of weighted social network, has on the voter model. The weights of the social links are drawn from a fixed distribution, independently of the degrees of the nodes. We also allow for antagonistic social links, such that individuals are influenced to take the opposite view of some neighbours \cite{li2015voter, li2013influence}. The weights of the social links therefore represent the degree of mutual respect (or disrespect if the link is antagonistic) that pairs of individuals have for one another's opinion.

The effect of a heterogeneously weighted interaction network is most evident in the biases of individual voters. Because of this, we focus primarily on the ensemble-averaged opinions of individuals rather than the population-averaged opinion, in contrast to many studies involving the voter model. A microscopic view allows us to distinguish between populations where individuals switch readily from one opinion to the other and more polarised populations where individuals are inclined to maintain their opinion.

It is deduced that the lifetimes of the biases of individual voters are characterised by the eigenvalues of the Laplacian matrix of the weighted social network. Laplacian matrices were also studied recently in the context of the Taylor model \cite{baumann2020laplacian}. Here, we make use of techniques which were originally developed for the study of dilute spin systems in order to evaluate the spectrum of eigenvalues of the network Laplacian. Using the effective-medium and single-defect approximations \cite{semerjian, birolisda}, we are able to produce closed-form expressions for the eigenvalue densities of a general class of sparse random matrices. This facilitates the systematic study of how the persistence of individual bias is affected by the various aspects of our model. 

We show that a sparse interaction network and a high degree of heterogeneity in the interaction strengths between individuals can give rise to long-lived biases for individual voters. We also demonstrate that a larger population of individuals increases the possibility of persistent bias. In the thermodynamic limit, it is possible to have arbitrarily long-lived individual biases even in circumstances where consensus is not reached.

The structure of the remainder of the manuscript is as follows: In Section \ref{sec:modeldef}, we define the modified voter model with quenched disorder and antagonistic interactions. Then in Section \ref{sec:relaxational}, we derive dynamical equations for the ensemble-averaged single-site biases. We also demonstrate that the relaxational dynamics of these quantities is dictated by the network Laplacian. Next, in order to provide a reference point for later results, we then take the simple example of a fully-connected, homogeneously-weighted network in Section \ref{sec:wellconnected}. In Section \ref{sec:theorygeneral}, we then give an overview of the analytical tools from the theory of sparse random matrices which we use to evaluate the eigenvalue spectrum of the interaction network Laplacian. In Sections \ref{sec:sdaer} and \ref{sec:generalvaryint} we then examine how the structure and size of the interaction network and the heterogeneity of the interaction strengths each affect the persistence of individual biases. Finally, we discuss the results and methods and conclude in Section \ref{sec:discussion}.

\section{Voter model with quenched disorder}\label{sec:modeldef}
In the classic network voter model \cite{castellano2009statistical, vazquez2008analytical, gleeson2013binary}, `voters' in a population are modelled as nodes on a network who each hold a binary opinion. Individual voters are selected at a constant rate to copy one of their neighbours. The individual to be copied is selected from amongst the neighbours of the primary voter with equal probability. If the neighbour and the primary voter have differing opinions, the primary voter is `persuaded' by the neighbour and changes its opinion.

We consider a set-up similar to this but with some modifications. More precisely, we imagine a set of $N$ individuals (or voters) indexed by $i$, each of whom has an opinion reflected by a binary state variable $s_i \in \{0,1\}$. Each individual interacts with a set of neighbours, causing it to possibly change its opinion. Neighbours may be connected by `reinforcing' or `antagonistic' links. A link (rather than a node) is chosen at a rate proportional to its weight. Then, if a `reinforcing' link is chosen, one of the adjacent nodes copies the other. If an `antagonistic' link is chosen, one node takes the opposite opinion to the other. 

The key differences between the model considered here and the classic network voter model are (a) the link-centred update procedure, (b) the weighted interaction network, (c) the possibility of antagonistic links. That is, the classic voter model and the model considered here are equivalent on a homogeneous complete graph with no antagonistic links. The link-centred update procedure was chosen in part for analytical convenience. We verify that our results are not changed qualitatively by using an alternative node-centred formulation in Appendix \ref{appendix:unweighted}. 

The aforementioned features of the model are encapsulated by the rates at which individuals adopt the opposite opinion. The rates at which a voter $i$ with opinion 0 switches to 1 ($r^+_i$) and 1 to 0 ($r_i^-$) are given by
\pagebreak 
\begin{align}
r^+_i &=  \left[ \sum_k  J^{(r)}_{ik}s_k  + \sum_k  J^{(a)}_{ik}(1-s_k) \right], \nonumber \\
r^-_i &=    \left[\sum_k J^{(r)}_{ik}(1-s_k) +  \sum_k J^{(a)}_{ik}s_k \right], \label{rates}
\end{align}
where $k$ sums over all nodes in the network, $J^{(r)}_{ik}$ are the coupling constants for pairs of spins which have a reinforcing link and $J^{(a)}_{ik}$ are the coupling constants for spins which have an antagonistic link. Defining the sub-network adjacency matrix for reinforcing links $\hat A$, we have 
\begin{align}
J^{(r)}_{ij} = A_{ij} J_{ij}, \,\,\,\,\, J^{(a)}_{ij} = (1-A_{ij}) J_{ij},
\end{align}
The matrices $\hat A$ and $\hat J$ are constrained to be symmetric. The elements of $\hat A$ are each selected to be $1$ with probability $1-f$ (a reinforcing link) and $0$ with probability $f$ (an antagonistic link). 

The constants $J_{ij}$ are drawn from a distribution such that they are guaranteed to be non-negative. This ensures the positivity of the rates in Eq.~(\ref{rates}). For analytical simplicity, we imagine that voters are connected on an Erd\"{o}s-R\'{e}nyi graph \cite{erdHos1960evolution}. That is, we choose $J_{ij}$ independently (with the condition $J_{ij} = J_{ji}$) from a probability distribution of the form
\begin{align}
P(J_{ij}) =\left(1-\frac{p}{N}\right) \delta(J_{ij}) + \frac{p}{N} \pi(J_{ij}), \label{sparsedef}
\end{align}  
where $\pi(J_{ij})$ is a probability distribution to be specified [satisfying $\pi(J_{ij}) = 0$ for $J_{ij}<0$], $\delta(\cdot)$ is the Dirac delta function and $p$ is the typical number of social connections per voter. One notes that by setting $p =  N$ with the choices $\pi(J_{ij}) = \delta(J_{ij}- 1/N)$ and $f = 0$, we recover the rates associated with the classic mean-field voter model from Eqs.~(\ref{rates}).

\section{Microscopic relaxational dynamics}\label{sec:relaxational}
Given the model set-up, we now derive dynamical equations for the biases of individual voters.  Consider the probability $P(\{ s_i\}, t)$ of observing a particular opinion configuration $\{ s_i\}$ at time $t$. This probability distribution obeys the following master equation \cite{lafuerza}
\begin{align}
\frac{d P(\{ s_i\}, t)}{dt} =& \sum_j (E_j - 1)[ s_j r_j^- P(\{ s_i\}, t)] \nonumber \\
& + \sum_j (E^{-1}_j - 1)[(1-s_j) r_j^+ P(\{ s_i\}, t)] ,
\end{align}
where the step operators take their usual meaning
\begin{align}
E_j f(s_1, \dots, s_j , \dots, s_N) &= f(s_1, \dots, s_j +1 , \dots, s_N), \nonumber \\
E_j^{-1} f(s_1, \dots, s_j , \dots, s_N) &= f(s_1, \dots, s_j -1 , \dots, s_N) .
\end{align}
One can thus deduce the following evolution equations for the first moments \cite{lafuerza}
\begin{align}
\frac{d \langle s_i \rangle}{dt } &= \langle r_i^+ \rangle - \langle (r^+_i + r_i^-) s_i \rangle ,\label{firstmoment}
\end{align}
where the angular brackets indicate an ensemble average over the intrinsic noise of the system for a single starting condition $\{s_i(0)\}$ and a fixed interaction network. Defining $m_i(t) = 2 \langle s_i(t) \rangle - 1$ for each site $i$, one obtains the following set of coupled ODEs from Eqs.~(\ref{rates}) and (\ref{firstmoment})  
\begin{align}
\frac{d m_i}{dt } =&   \sum_k (2 A_{ik} -1)J_{ik} m_k -  m_i\sum_k J_{ik} . \label{momentevol}
\end{align}
Conveniently, Eq.~(\ref{momentevol}) doesn't depend on any higher moments of $s_i(t)$. For $f = 0$, Eq.~(\ref{momentevol}) reduces to a network diffusion equation (which has previously been derived for the standard network voter model \cite{castellano2009statistical}). 

The set $\{m_i(t)\}$ are the quantities central to our consideration. They can be thought of as the statistical `biases' that each individual has for one opinion over the other. An alternative set of quantities $\{ C_i(t,0)\}$, the correlations that each opinion has with its starting value, is discussed in Appendix \ref{appendix:correlations}. 

Due to the linearity of Eq.~(\ref{momentevol}), the time evolutions of the average biases $\{m_i(t)\}$ are characterised by the eigenvalues of the matrix $\hat M$ with entries
\begin{align}
M_{ij} = (2 A_{ij} -1)J_{ij}  - \delta_{ij} \sum_k J_{ik} , \label{matdef}
\end{align}
where $\delta_{ij}$ is the Kronecker delta. Since $\{m_i(t)\}$ are bounded such that $-1 \leq m_i \leq 1$, $\hat M$ cannot possibly have any positive eigenvalues. However, it can have eigenvalues equal to zero and arbitrarily small negative eigenvalues. The existence of such eigenvalues would imply a long-lasting bias for voters. Using replica techniques to deduce the eigenvalue spectra of matrices of the form $\hat M$, we are able to determine the factors that contribute to long-lasting biases. 

\begin{widetext}
	
	\begin{figure}[h]
		\centering 
		\includegraphics[scale = 0.4]{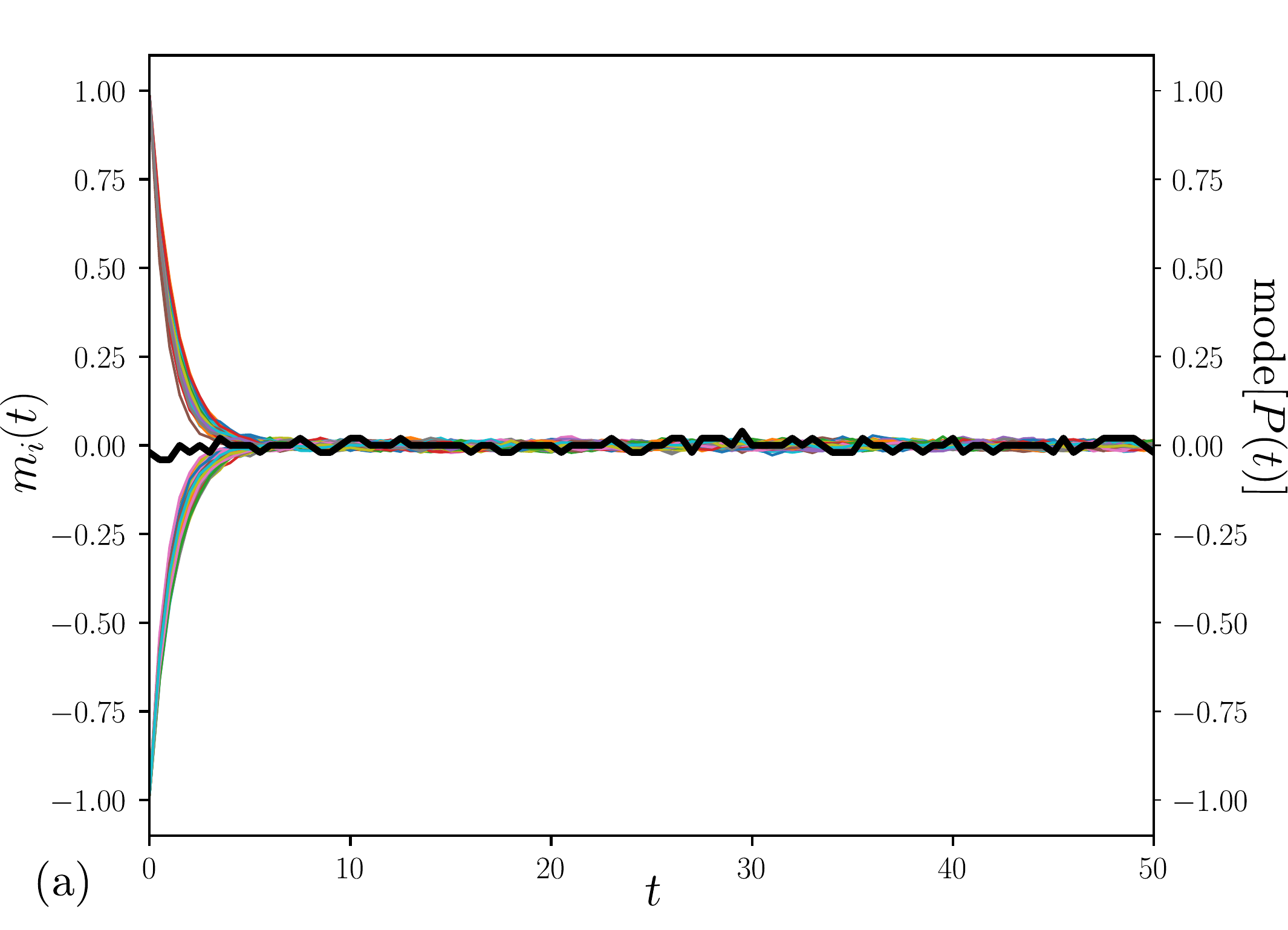}
		\includegraphics[scale = 0.4]{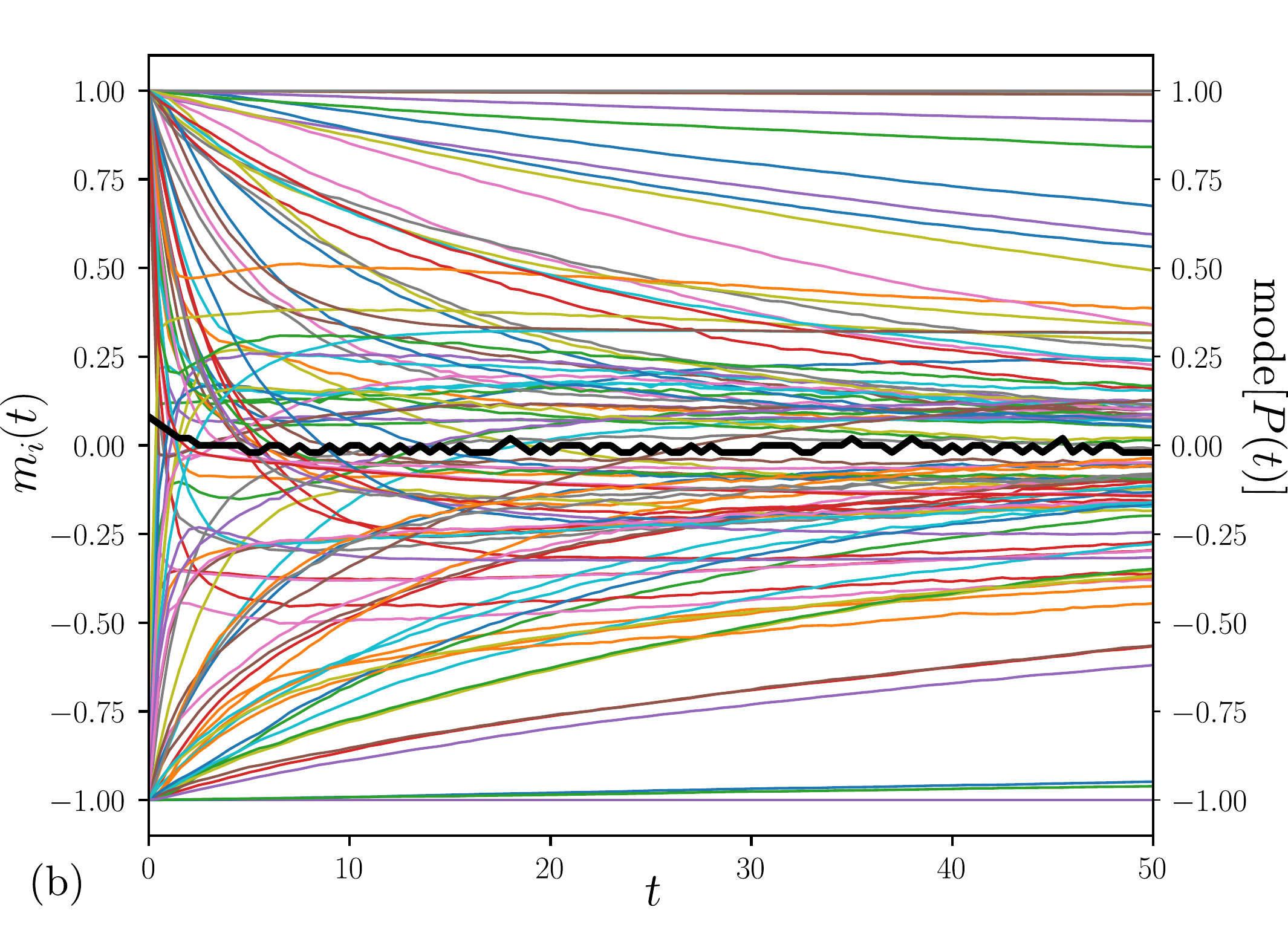}
		\caption{Results of individual-based simulations of the dynamics determined by the rates in Eq.~(\ref{rates}) averaged over $20000$ trials for the same starting conditions. We plot both the individual biases $m_i(t)$ (thin coloured lines) and the mode of the distribution of the macroscopic order parameter $P(t) = 2 N^{-1}\sum_{i} s_i(t) - 1$ (thick black line). The presence of antagonistic links ($f>0$) means that consensus ($P = \pm1$) cannot occur and $P(t)$ instead fluctuates around a value of $0$ indefinitely. In panel (a), each individual is equally likely to adopt either opinion after an initial transient period. In panel (b), individuals maintain a long-lived bias for one opinion over the other. In both panels, we use $N = 100$ and $f = 0.5$. In panel (a), $\pi(J)$ is as defined in Eq.~(\ref{simplepi}) so that each non-zero off-diagonal element of $\hat M$ has the same magnitude and $p = 100$. In panel (b), $\pi(J)$ is gamma distributed [see Eq.~(\ref{gammadist})], with $p=10$, $m =1$ and $s = 2$.  } \label{fig:individualbased}
	\end{figure}
\end{widetext}

\section{Special case: complete graph}\label{sec:wellconnected}
For the sake of developing some intuition and in order to contrast against later results, we discuss the behaviour of the model when the social network is a complete graph. We first consider the dynamics of the quantities $\{m_i\}$ when there are no antagonistic links ($f=0$) and then discuss the effect of introducing such links ($f>0$). For now, we restrict ourselves to a homogeneous network such that $\pi(J_{ij}) = \delta(J_{ij} - 1/N)$.

\subsection{No antagonistic links ($f = 0$)}
We first examine the case of the classic voter model by setting $f = 0$, $\pi(J_{ij}) = \delta(J_{ij}- 1/N)$ and $p = N$. The matrix $\hat M$ in this case always has an eigenvector $v^{(0)} =[1,1, \cdots, 1]$ with corresponding eigenvalue $\mu = 0$. The remaining $N-1$ eigenvectors are all orthogonal to this (following from the symmetry of $\hat M$) and have degenerate eigenvalues at $\mu = -1$. If we decompose the initial single-site averages $m_i(0)$ in terms of the eigen-basis of $\hat M$ so that $m_i(0) = \sum_{\mu} c_{(\mu)} v^{(\mu)}_i$, one finds
\begin{align}
m_i(t) = \sum_{\mu} c_{(\mu)} v^{(\mu)}_i e^{-\mu t} . \label{eigenbasisdecomp}
\end{align}
As $t \to \infty$, all components with non-zero eigenvalues vanish and we are left with $m_i(\infty) = c_{(0)} v^{(0)}_i = c_{(0)}$. Since all other eigenvectors are orthogonal to $v^{(0)}$, we must have that $c_{(0)} = N^{-1}\sum_i m_{i}(0)$. That is, in the long-run, each single-site bias tends to the population-averaged initial bias. 

This is connected to the fact that, in the classic voter model, consensus is always eventually achieved for finite $N$. Consensus is achieved at $\sum_i s_i(\infty) = N$ with probability $N^{-1}\sum s_i(0)$ and at $\sum_i s_i(\infty) =0$ with probability $N^{-1}\sum [1-s_i(0)]$. So, because consensus is always reached for finite $N$, there is always a persistent single-site bias.

\subsection{Effect of antagonistic links ($f>0$)} \label{sec:effectofantagonistic}
Let us now consider the case where $f \neq 0$ and $\pi(J_{ij}) = \delta(J_{ij}- 1/N)$ and $p = N$. Now, no consensus is reached in the limit $t \to \infty$. Rather like the case with contrarian voters \cite{masuda2013, khalil2019noisy}, the presence of antagonistic connections between voters leads to a frustrated dynamical state. This effect is demonstrated in Fig.~\ref{fig:individualbased}a. 

Considering now the microscopic dynamics, we see that (following heuristic arguments along the lines of Ref.~\cite{allesinatang2}) $v^{(0)}$ remains an eigenvector of the matrix $\hat M$ for large $N$, but now the corresponding eigenvalue is given by $\mu = - 2f$. The remaining eigenvalues are again clustered around $\mu = - 1$ \cite{wigner1958distribution}. So in this case, each voter eventually arrives at a bias of $m_i = 0$. That is, the initial condition no longer has an influence over the long-term dynamics (as is also demonstrated in Fig. \ref{fig:individualbased}a).

In contrast to this, we demonstrate in the following sections that the combination of a sparse interaction network and variation in the random matrix elements $J_{ij}$ can lead to a long-lasting bias for individual voters even when $f \neq 0$. This means that although consensus is not achieved, individual voters may still retain their biases for an arbitrarily long time. This is exemplified in Fig.~\ref{fig:individualbased}b. Using techniques from the study of sparse random matrices, we are able to systematically study which factors give rise to persistent bias.

\vspace{4pt}

\section{Approximating the eigenvalue spectrum of the matrix $\hat M$}\label{sec:theorygeneral}
In order to characterise the lifetimes of the individual biases $m_i$ under more general conditions, we must develop a broader understanding of the spectrum of eigenvalues of $\hat M$. 

The random matrix $\hat M$ defined in Eq.~(\ref{matdef}) has a finite average number of non-zero entries per row/column $p$. This number is taken to be an independent parameter from the total number of rows/columns $N$. In particular, when the limit $N \to \infty$ is taken (a useful consideration when approximating the eigenvalue spectrum), $p$ remains finite. Such random matrices are known as sparse \cite{kuhn2008spectra, metz2019spectral}. The sparse nature of $\hat M$, combined with the dependence of its diagonal entries on other elements, means that its eigenvalue spectrum deviates substantially from the prototypical Wigner semi-circle law \cite{wigner1958distribution}.

A number of analytical methods exist for the purpose of evaluating the eigenvalue spectra of sparse random matrices \cite{kuhn2008spectra, metz2019spectral, rodgersbray, semerjian, birolisda}. Here we generalise the Effective-Medium (EMA) and Single-Defect (SDA) approximations, which have so far largely only been used to evaluate the eigenvalue spectra of random matrices with non-zero elements with fixed magnitude \cite{semerjian, birolisda, nagao2007spectral}. The advantage of these approximation schemes over others is that they yield closed-form expressions for the eigenvalue density, which we can then use to efficiently estimate the expected leading eigenvalue. Crucially, we obtain expressions which hold in the thermodynamic limit $N \to \infty$.

A more detailed overview of the origin of the EMA and the SDA is given in Appendix \ref{appendix:summary}. In summary, the EMA approximation $\rho^{\mathrm{EMA}}(\mu)$ to the eigenvalue density of $\hat M$ is given by (see Appendix \ref{appendix:emasummary})
\begin{align}
\rho^{\mathrm{EMA}}(\mu) &= - \frac{1}{\pi} \mathrm{Im} \, \sigma(\mu + i\epsilon) , \nonumber \\
\mu(\sigma) &= \frac{1}{\sigma} - \frac{ p}{2\sigma} - \frac{ p}{4\sigma^2} \int dJ \,     \frac{ \pi(J)}{\frac{1}{2\sigma}+ J} ,  \label{emasummary}
\end{align}
where $\epsilon$ is a small, real regulariser which is introduced to avoid divergences \cite{sommers, kuhn2008spectra}. One solves the second of Eqs.~(\ref{emasummary}) for $\sigma(\mu)$ and substitutes this into the first equation to obtain the EMA spectrum. To the Author's knowledge, this result has not been written explicitly in this form in previous literature. Conveniently, we see that the object on the right-hand side of this equation is related to the Stieltjes transform \cite{bateman1954tables} of the probability distribution $\pi(J)$. This observation allows one to exploit tables of integral transforms when attempting to solve for $\sigma(\mu)$.

Although the EMA replicates the main `bulk' of the eigenvalue distribution fairly well, it does a poor job of recovering the tails of the distribution, which are associated with localisation effects \cite{birolisda, Bray_1982}. For this reason, it is beneficial to go one level of approximation further.

As is derived in Appendix \ref{appendix:sdasummary}, the more accurate but less analytically manageable, SDA is given by 
\begin{widetext}
	
\begin{align}
\rho^{\mathrm{SDA}}(\mu) &=  \sum_{k=0}^\infty \frac{e^{-p} p^k}{k!} \left(-  \frac{1}{\pi}\right)\mathrm{Im} \left\{ \int \prod_{r = 1}^k \left[d J_r \pi(J_r) \right]\frac{1}{\mu + i \epsilon + \sum_{r=1}^k [1/J_r+ \sigma(\mu + i \epsilon)]^{-1}}\right\} , \label{sdasummary}
\end{align}
\end{widetext}
where $\sigma(\mu)$ is again given by inverting the second of Eqs.~(\ref{emasummary}). Although this expression is analytically unwieldy in its full generality, it can be used to give important insight into the eigenvalue spectrum and can be simplified significantly in certain special cases, as we shall demonstrate. Examples comparing the EMA and SDA to numerically evaluated eigenvalue spectra are given in Figs. \ref{fig:emavssdaer} and \ref{fig:gamma}.

\section{The effect of a sparse interaction network on individual bias}\label{sec:sdaer}
Having derived general expressions for the eigenvalue spectrum of the matrix central to our problem $\hat M$, we are now able to characterise the relaxational dynamics in some enlightening special cases. We first examine the simple case where 
\begin{align}
\pi(J_{ij}) = \delta\left(J_{ij} - \frac{1}{p}\right). \label{simplepi}
\end{align}
Studying this special case allows us to isolate the effects of a sparse random graph on the relaxational dynamics. We study the effect of a variation in link strength later in Section \ref{sec:generalvaryint}. One notes that by choosing $A_{ij} = 1$ for all pairs $(i,j)$ in combination with Eq.~(\ref{simplepi}), the matrix $\hat M$ defined in Eq.~(\ref{matdef}) becomes the Laplacian of the Erd\"{o}s-R\'{e}nyi random graph \cite{birolisda, brayrodgers}.  

Due to the variation in the number of non-zero entries from row to row, the eigenvalues of $\hat M$ are no longer as simply distributed as in the all-to-all case discussed in Section \ref{sec:effectofantagonistic}. Instead of the majority of eigenvalues clustering closely around $\mu = -1$, the eigenvalues become more broadly distributed as $p$ is reduced. 

In the following subsections, we first present analytical expressions for the eigenvalue spectrum of $\hat M$ in this special case. We then go on to use these results to deduce the behaviour of the expected leading eigenvalue as a function of $p$ and $N$.

\subsection{Outlier eigenvalue due to antagonistic links}\label{sec:outlierantagonistic}
On the macroscopic level, the presence of antagonistic links ($f>0$) prevents the system from reaching an absorbing state, as discussed in Section \ref{sec:effectofantagonistic}. The dynamics continue indefinitely in a frustrated manner (see Fig. \ref{fig:individualbased}). However, the bulk of the eigenvalue spectrum of $M_{ij}$ is actually independent of the fraction of antagonistic links $f$. This is shown in Appendix \ref{appendix:emasummary} and is evident from Eq.~(\ref{emasummary}). That is, the bulk of the eigenvalue spectrum of $\hat M$ is equivalent to that of the Laplacian of an Erd\"{o}s-R\'{e}nyi random graph \cite{birolisda} for all values of $f$ when Eq.~(\ref{simplepi}) is satisfied . 

There is however a single additional eigenvalue which is missed by both the EMA and SDA (as discussed in Section \ref{sec:wellconnected}) at approximately $-2f$ (see Fig. \ref{fig:emavssdaer}). The only effect of varying $f$ on the microscopic dynamics is the location of this single eigenvalue \cite{semerjian}. When $f \gtrapprox 0.5$, this additional outlier is not relevant to the persistence of bias. 

\subsection{Bulk eigenvalue density and Lifshitz tail of the Erd\"{o}s-R\'{e}nyi graph Laplacian}
The eigenvalue spectrum of the Laplacian of an Erd\"{o}s-R\'enyi graph was studied using the SDA in \cite{birolisda}. However, only numerical results were presented. Here, we present closed-form expressions for the EMA and SDA in this special case.

The EMA for the eigenvalue density (see Appendix \ref{appendix:emasummary}) is given by
\begin{align}
\rho^{\mathrm{EMA}}(\mu) = -\frac{\sqrt{-[p(1 + \mu)-2]^2-8 \mu p}}{4 \pi \mu} . \label{emalaplacian}
\end{align}
According to this expression, the bounds of the eigenvalue spectrum are given by 
\begin{align}
\mu_{\pm} = -\left(1\pm\sqrt{\frac{2}{p}}\right)^2. \label{boundsema}
\end{align} 
The SDA can be used to improve upon the accuracy of Eq.~(\ref{emalaplacian}) and to identify the locations of outlier eigenvalues which are missed by the EMA. The SDA in this case is given by
\begin{align}
\sigma(\mu) &= \frac{2 - (\mu + 1) p - \sqrt{8 \mu p + (p(\mu+1) -2)^2}}{4\mu}, \nonumber \\
\rho^{\mathrm{SDA}}(\mu) &=  \sum_{k=0}^\infty \frac{e^{-p} p^k}{k!} \left( - \frac{1}{\pi}\right)\mathrm{Im} \left[\frac{1}{\mu + i \epsilon+ \frac{k}{p+\sigma(\mu + i \epsilon)}}\right],  \label{sdalaplacian}
\end{align} 
where again $\epsilon$ is a small real number. According to Eq.~(\ref{sdalaplacian}), outside the region where the EMA eigenvalue density is non-zero, one finds Dirac-delta peaks at locations which satisfy $\mu_k[p +\sigma(\mu_k)] = k$ for $k = 0, 1, 2, \cdots$. Or, more explicitly, 
\begin{align}
\mu_k = \frac{-p - 3 k p +p^2 - p \sqrt{(k-1)^2 +(p-1)^2  - 2 k p -1}}{2 p^2}. \label{deltapositions}
\end{align}
We note the negative sign of the square root in the above expression ensures that these eigenvalues are negative. The weights of these delta peaks are given by 
\begin{align}
w_k = e^{-p} \frac{p^k}{k!} \left[1 - \frac{ \mu_k^2\sigma'(\mu_k)}{k^2}\right]^{-1} , \label{weights}
\end{align}
where $\sigma'(\mu) = d\sigma/d\mu$ and $\sigma(\mu)$ is given by the first of Eqs.~(\ref{sdalaplacian}). 

\begin{widetext}
	
\begin{figure}[H]
	\centering 
	\includegraphics[scale = 0.5]{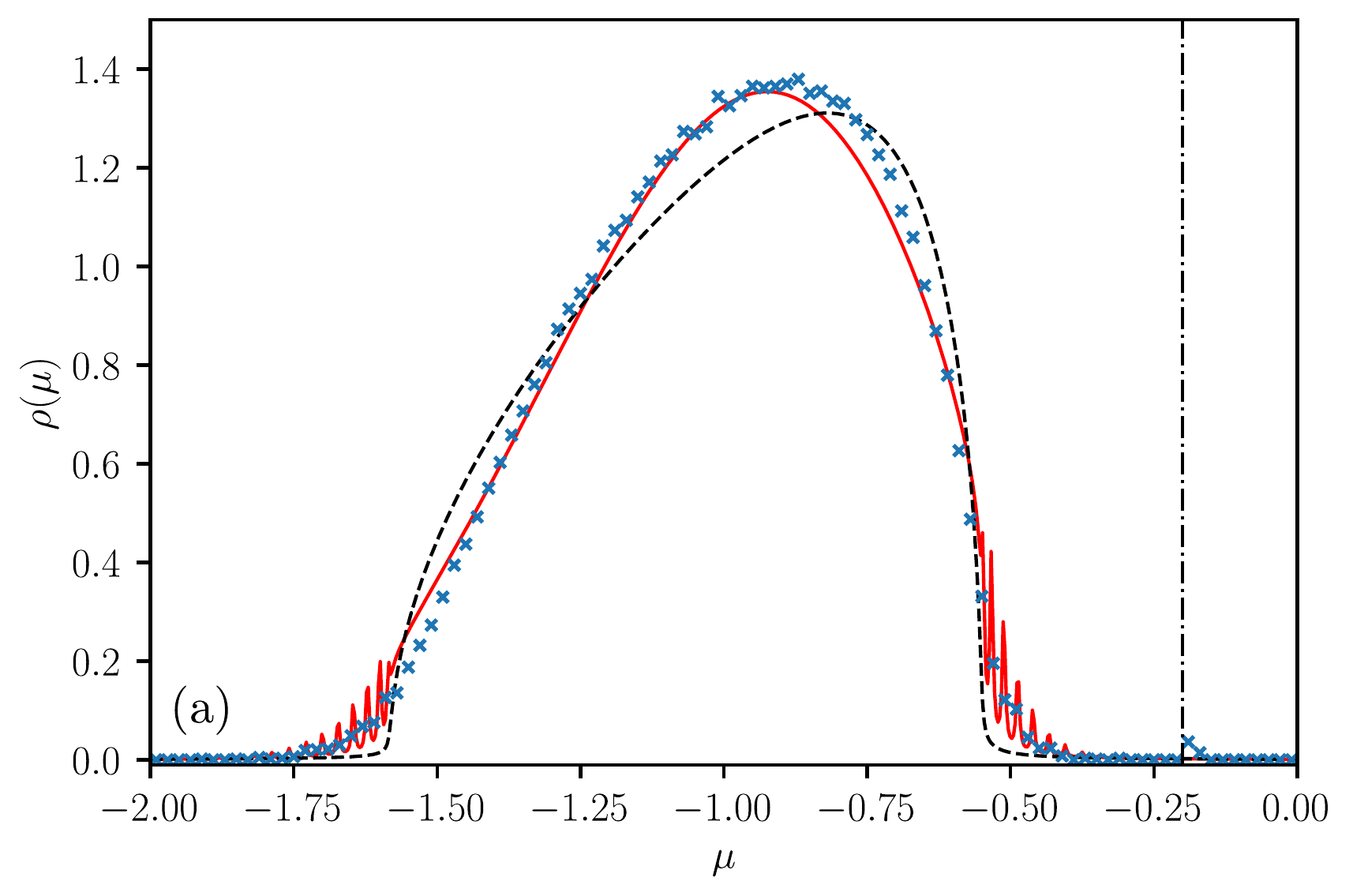}
	\includegraphics[scale = 0.5]{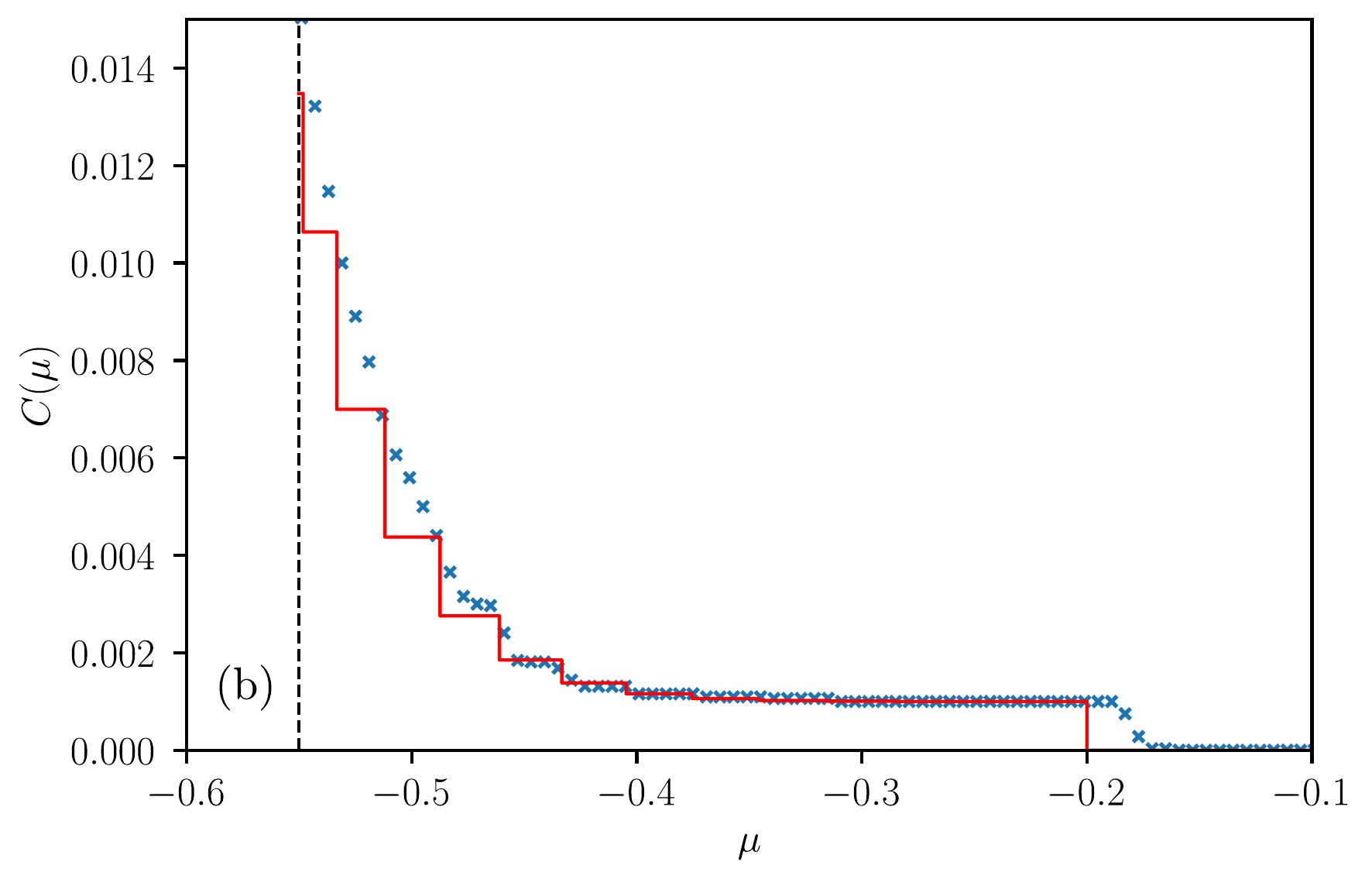}
	\caption{Panel (a): Comparison of the EMA [dashed black line -- see Eq.~(\ref{emalaplacian})] and the SDA [solid red line -- see Eq.~(\ref{sdalaplacian})] with regulariser $\epsilon = 0.003$ to the results of numerical diagonalisation. Using a non-zero value of $\epsilon$ smears out the Dirac-delta peaks in the SDA, making them visible for the sake of comparison to numerical results. The dot-dashed vertical black line indicates the heuristic prediction for the delta-peak of magnitude $1/N$ which arises due to the presence of antagonistic links at $\mu_f \approx -2f$ (see Section \ref{sec:outlierantagonistic}). Panel~(b): The integrated eigenvalue density, verifying the weights and positions of the peaks outside the bulk of the spectrum [see Eqs.~(\ref{deltapositions}) and (\ref{weights})]. The regulariser $\epsilon$ is now set to zero for the red theory line. The dashed vertical line indicates the position of the edge of the bulk region of the spectrum, as predicted by the EMA [see Eq.~(\ref{boundsema})]. In both panels, $\pi(J_{ij}) = \delta(J_{ij} - 1/p)$, $f = 0.1$, $p = 30$ and $N= 1000$. The blue crosses are the numerical results, averaged over 32 realisations of the matrix entries. } \label{fig:emavssdaer}
\end{figure}
\end{widetext}

The EMA and the SDA predictions for the eigenvalue spectrum are compared to the results of numerical diagonalisation in Fig.~\ref{fig:emavssdaer}a. Indeed, the SDA provides a more accurate approximation of the bulk region than the EMA. The weights of the delta peaks predicted by the SDA are verified in Fig.~\ref{fig:emavssdaer}b using the integrated eigenvalue density $C(\mu) = \int_\mu^0 d\lambda \,\rho(\lambda)$. 

One notes that in the numerical results the delta peaks are `smeared' across a range of values of $\mu$, creating a smooth ``Lifshitz tail'' \cite{Lifshitz_1965,khorunzhiy2006lifshitz,bray1987nature, rodgersmoore1989distribution,Bray_1982}. The integrated eigenvalue density therefore varies continuously with $\mu$ for the numerical results, in contrast to the theoretical prediction which jumps discontinuously. Despite this smearing effect however, the magnitudes of the peaks are well-predicted by the SDA [see Eq.~(\ref{weights})]. The smearing effect appears not to be alleviated by using a larger matrix size $N$.

\subsection{Expected leading eigenvalue: dependence on $p$ and $N$}
The typical relaxation time of the system is characterised by the leading (rightmost) eigenvalue of the matrix $\hat M$. Having verified the EMA and SDA, we now use these approximations to evaluate the expected leading eigenvalue (ELE) and examine its dependence on the parameters $p$ and $N$.

There exists a broad literature on the subject of leading eigenvalues of random matrices and for adjacency matrices in particular (see for example \cite{restrepo2007approximating, krivelevich2003largest, hong2005sharp}). Because the leading eigenvalues cluster in a series Dirac-delta peaks in our case, it is possible for us to estimate the ELE analytically for large $N$ using the SDA in a relatively simple fashion.

For the present discussion, we imagine that $f$ is sufficiently large so that the single eigenvalue associated with a non-zero $f$ (denoted $\mu_f$) is such that $\mu_f<\mu_-$ [see Eq.~(\ref{boundsema})]. We also approximate each eigenvalue of the random matrix as being drawn independently from the distribution $\rho^\mathrm{SDA}(\mu)$ [see Eq.~(\ref{sdalaplacian})]. 

The probability that a particular eigenvalue of $\hat M$ takes the value $\mu_k$ is $w_k$ [given by Eq.~(\ref{weights})]. So the probability that none of the eigenvalues of $\hat M$ take the value $\mu_1$ is $q_1 = (1-w_1)^N$. This means that the probability that $\mu_1$ is the largest eigenvalue is given by
\begin{align}
p_1 = 1- (1-w_1)^N.
\end{align}
The probability that the largest eigenvalue is $\mu_2$ is then $p_2 = q_1 [1-(1-w_2)^N]$. The pattern continues so that the probability that $\mu_k$ is the largest eigenvalue is 
\begin{align}
p_k = [1-(1-w_k)^N] \prod_{l = 1}^{k-1} (1-w_l)^N .
\end{align}
We make the further approximation that if none of the Dirac-delta peaks attract any eigenvalues, then the leading eigenvalue is given by the edge of the bulk of the eigenvalue spectrum as predicted by the EMA [see Eq.~(\ref{boundsema})]. That is, the probability that the largest eigenvalue is $\mu_-$ is given by 
\begin{align}
	p_- = \prod_{l = 1}^{k_{\mathrm{max}}} (1-w_l)^N,
\end{align}
where $k_{\mathrm{max}}$ is the number of Dirac-delta peaks with $\mu_k~>~\mu_-$. One can verify $p_- + \sum_{l = 1}^{k_{\mathrm{max}}} p_l =1$. The expected leading eigenvalue is then given by
\begin{align}
\left\langle\mu_{\mathrm{max}}\right\rangle_{M} = p_- \mu_- + \sum_{k=1}^{k_\mathrm{max}} p_k \mu_k , \label{expectedlargesteigenvalue}
\end{align}
where the angular brackets $\langle \cdot \rangle_M$ represent an average over realisations of the random matrix $\hat M$. 

\begin{figure}[H]
	\centering 
	\includegraphics[scale = 0.5]{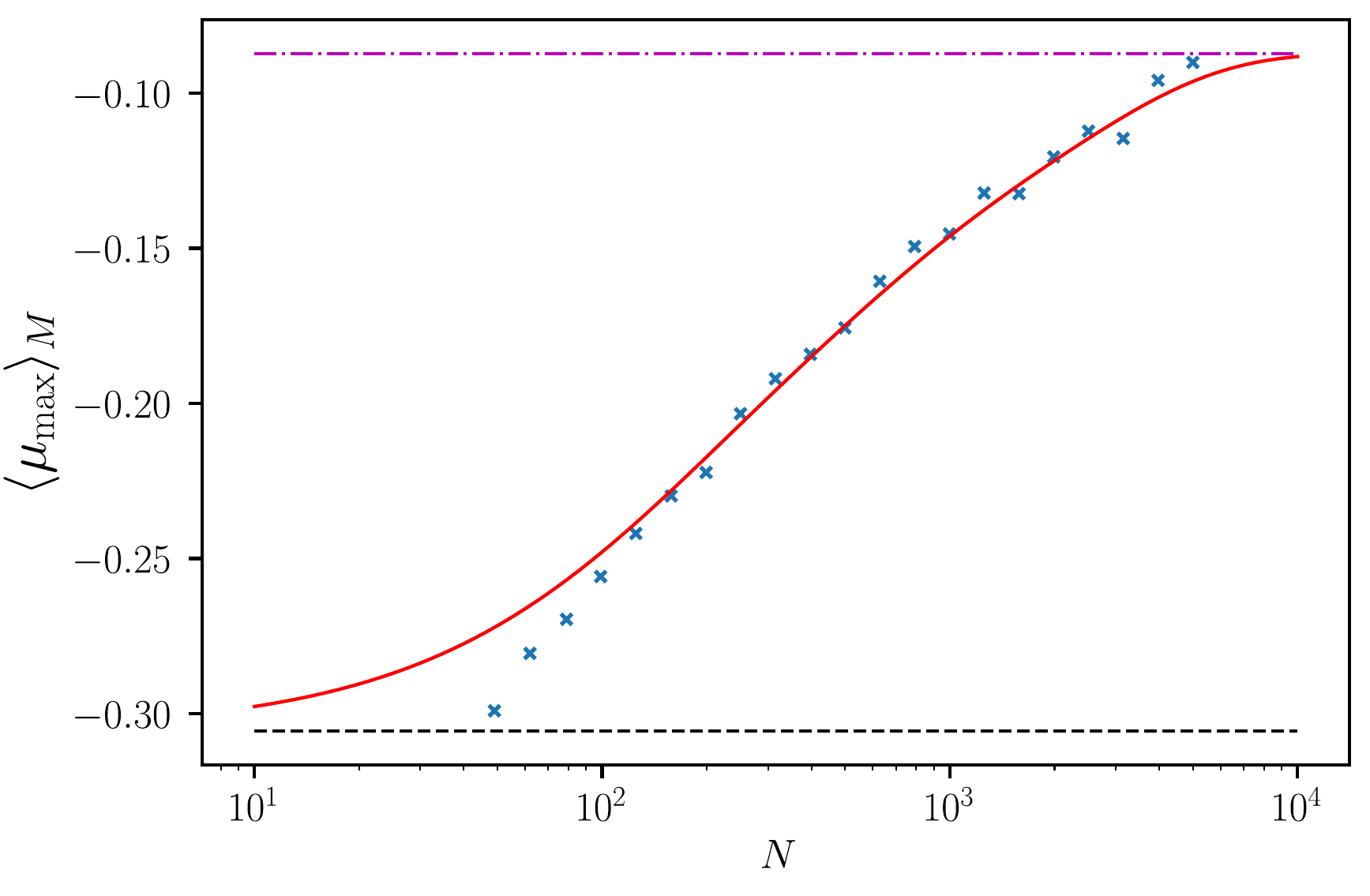}
	\caption{The expected largest eigenvalue of the model with $\pi(J) = \delta(J-1/p)$ as a function of the total number of voters $N$. Here $p = 10$ and $f = 0.5$ and the numerical results were averaged over 32 trials. The horizontal dot-dashed magenta line at the top of the figure is the position of the largest delta peak at $\mu_1$ [see Eq.~(\ref{deltapositions})]. The horizontal dashed black line at the bottom of the figure is the edge of the bulk of the eigenvalue spectrum $\mu_-$ [see Eq.~(\ref{boundsema})]. The solid red curve is the theoretical prediction in Eq.~(\ref{expectedlargesteigenvalue}). } \label{fig:mumaxversusn}
\end{figure}

From Eq.~(\ref{expectedlargesteigenvalue}), we can extract the $N$-dependence of the expected leading eigenvalue. This is shown in Fig.~\ref{fig:mumaxversusn} along with the results of numerical diagonalisation. We note that the ELE deviates more significantly from the edge of the EMA bulk spectrum the more $N$ is increased. It thus becomes more necessary to take into account the tails of the eigenvalue spectrum for large $N$. In the limit $N \to \infty$, it is almost certain that one eigenvalue will occupy the value $\mu_1$, so we see $\left\langle\mu_{\mathrm{max}}\right\rangle_M \to \mu_1$ as $N\to \infty$. Since our theoretical approach is based on a saddle-point approximation, which assumes large $N$, we should not expect our results to hold for smaller values of $N$. 
	
\begin{figure}[h]
	\centering 
	\includegraphics[scale = 0.5]{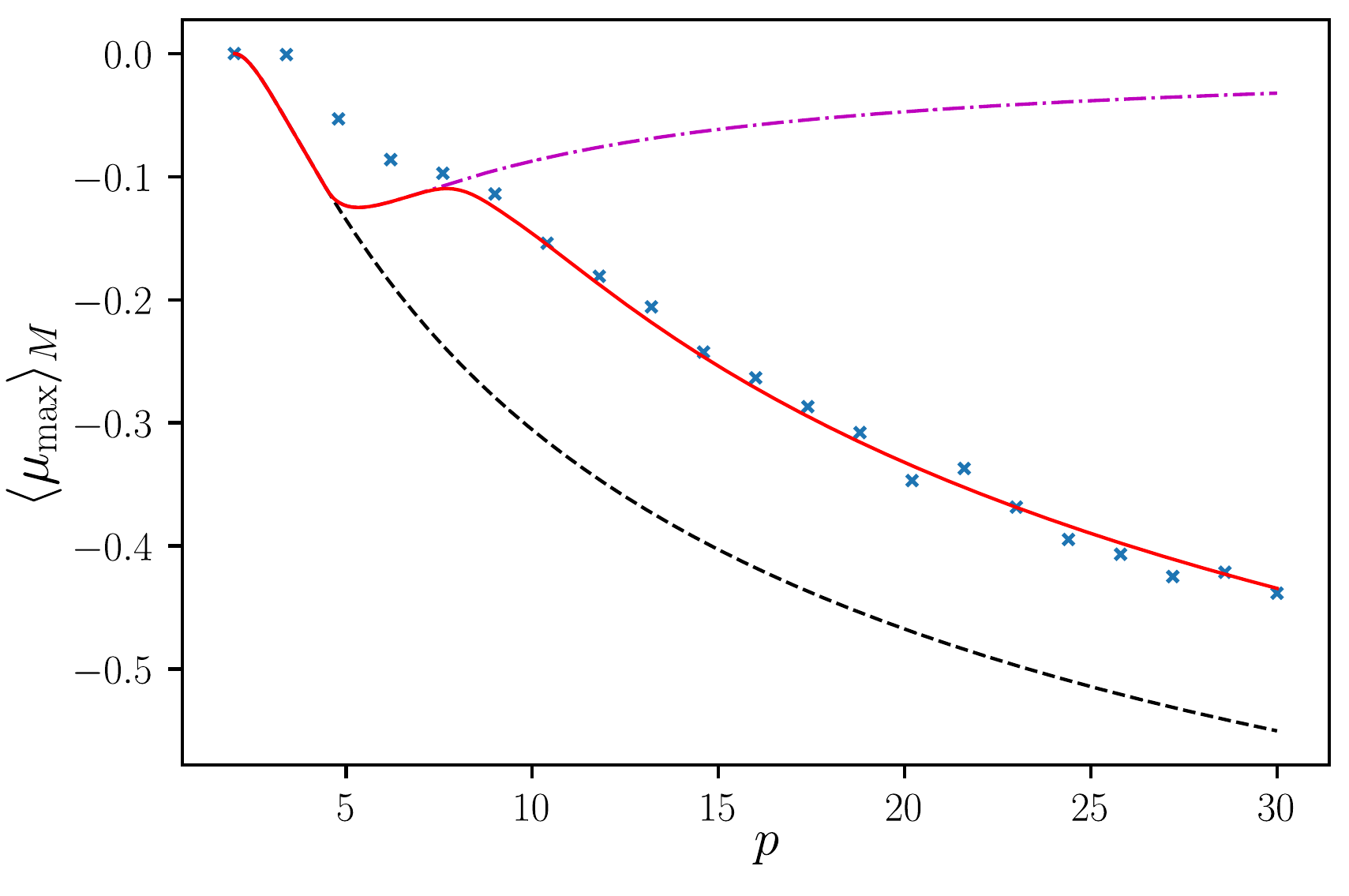}
	\caption{The expected largest eigenvalue of the model with $\pi(J) = \delta(J-1/p)$ as a function of the total number of voters $N$. Here $N = 1000$ and $f = 0.5$ and the numerical results were averaged over 32 trials. The three theory lines are as described in the caption of Fig. \ref{fig:mumaxversusn}. } \label{fig:mumaxversusp}
\end{figure}

We also demonstrate the behaviour of the ELE as a function of $p$ in Fig. \ref{fig:mumaxversusp}. The SDA captures the behaviour well for moderate to large values of $p$. We see that as $p$ is increased, the eigenvalue spectrum becomes more compact. For $p \to \infty$, the distribution tends towards and Dirac delta peak centred on $\mu = -1$ (as discussed in Section \ref{sec:wellconnected}). 

On the other hand, the approximations fail to capture the precise behaviour for small $p$. This is because the Gaussian approximation involved in the EMA (see Appendix \ref{appendix:emasummary}) is only accurate for large $p$ \cite{semerjian}. Despite this, the prediction of the ELE approaching $\mu = 0$ for $p = 2$ [see Eq.~(\ref{boundsema})] may have some significance. For regular lattices with dimension $d\leq 2$, consensus is always reached in the classic voter model even in the thermodynamic limit \cite{castellano2009statistical}. 

As we discuss in Section \ref{sec:generalvaryint}, it is possible for this threshold value of $p$ to increase from 2 when variation in link strength is introduced, as is shown in Fig. \ref{fig:mumaxversuss}. We note that the predicted threshold value is thus always above the percolation threshold for Erd\"{o}s-R\'{e}nyi graphs at $p =1$ \cite{semerjian, bollobas1998random}.

\section{The effect of distributed interaction strength on individual bias}\label{sec:generalvaryint}

Having examined the simple case where all non-zero elements of $J_{ij}$ take the same value, we now consider an example where the magnitudes of the non-zero elements may take any value over the positive real axis. Consider the gamma distribution
\begin{align}
\pi(J) = \frac{1}{\Gamma(k) \theta^k} J^{k-1} e^{-\frac{J}{\theta}} . \label{gammadist}
\end{align}
We introduce the parameters $m$ and $s^2$ via 
\begin{align}
\langle J \rangle = m/p , \,\,\,\, \langle (J - \langle J \rangle)^2\rangle = s^2/p. \label{defms}
\end{align}
The choice of scaling with $p$ in Eq.~(\ref{defms}) is common and ensures sensible behaviour for large $p$ \cite{sommers, edwardsjones}. Noting that for the gamma distribution we have $\langle J \rangle = k \theta$ and $\langle (J - \langle J \rangle)^2\rangle = k \theta^2$, Eq.~(\ref{defms}) requires that $k = \frac{m^2}{s^2 p}$ and $\theta = \frac{s^2}{m}$. 

\subsection{Bulk of the eigenvalue spectrum}
\begin{widetext}
	
	\begin{figure}[H]
		\centering 
		\includegraphics[scale = 0.37]{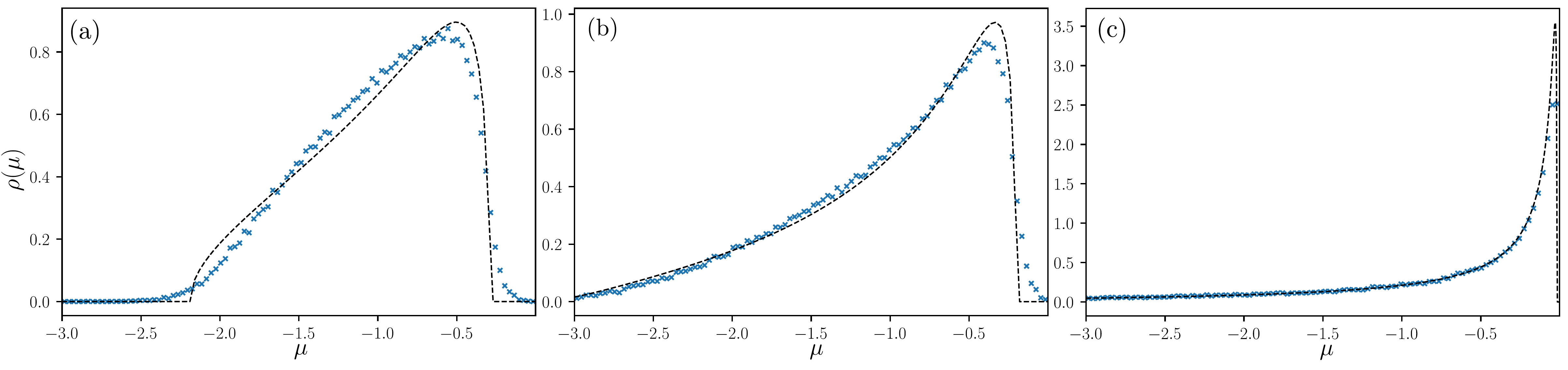}
		\caption{ Verification of the EMA when the non-zero matrix elements are gamma-distributed for various values of the variance. In Panel (a) $s = 0.01$, in panel (b) $s = 0.1$ and in panel (c) $s = 1$. As the variance of the non-zero random matrix elements is increased and the mean is held constant, the right-hand tail of the eigenvalue distribution begins to extend towards zero. The blue crosses are the results of numerical diagonalisation for $m=1$, $N = 100$, $f = 0.5$ and $p = 10$ averaged over 1000 realisations of the matrix entries. The dashed black line is the EMA prediction from Eq.~(\ref{emageneral}). } \label{fig:gamma}
	\end{figure}
\end{widetext}

In order to evaluate the EMA of the eigenvalue spectrum of the matrix $\hat M$ according to Eqs.~(\ref{emasummary}), the Stieljes transform of the distribution $\pi(J)$ is required. The Stieltjes transform of the gamma distribution in Eq.~(\ref{gammadist}) is known and is given by \cite{bateman1954tables}
\begin{align}
\int dx \frac{\pi(x)}{x + y } = \frac{ y^{k-1} }{\theta^k} e^{\frac{y}{\theta}} \Gamma(1-k, \frac{y}{\theta}).
\end{align}
We thus arrive at a closed-form expression for EMA in this case
\begin{align}
\mu(\sigma) =& \frac{1}{\sigma} - \frac{ p}{2\sigma} \nonumber \\
&+ \frac{p (2\sigma)^{1-k}}{4 \sigma^2 \theta^k} \exp\left(\frac{1}{2\theta \sigma }\right) \Gamma\left(1-k, \frac{1}{2\theta \sigma}\right)  . \label{emageneral}
\end{align}
Eq.~(\ref{emageneral}) can be solved to yield $\sigma(\mu)$, which can in turn be used to yield the eigenvalue density via the first of Eqs.~(\ref{emasummary}). The results of doing so are compared with computationally generated eigenvalue spectra in Fig. \ref{fig:gamma}.

\subsection{Expected leading eigenvalue: dependence on $s$}

\begin{figure}[h]
	\centering 
	\includegraphics[scale = 0.5]{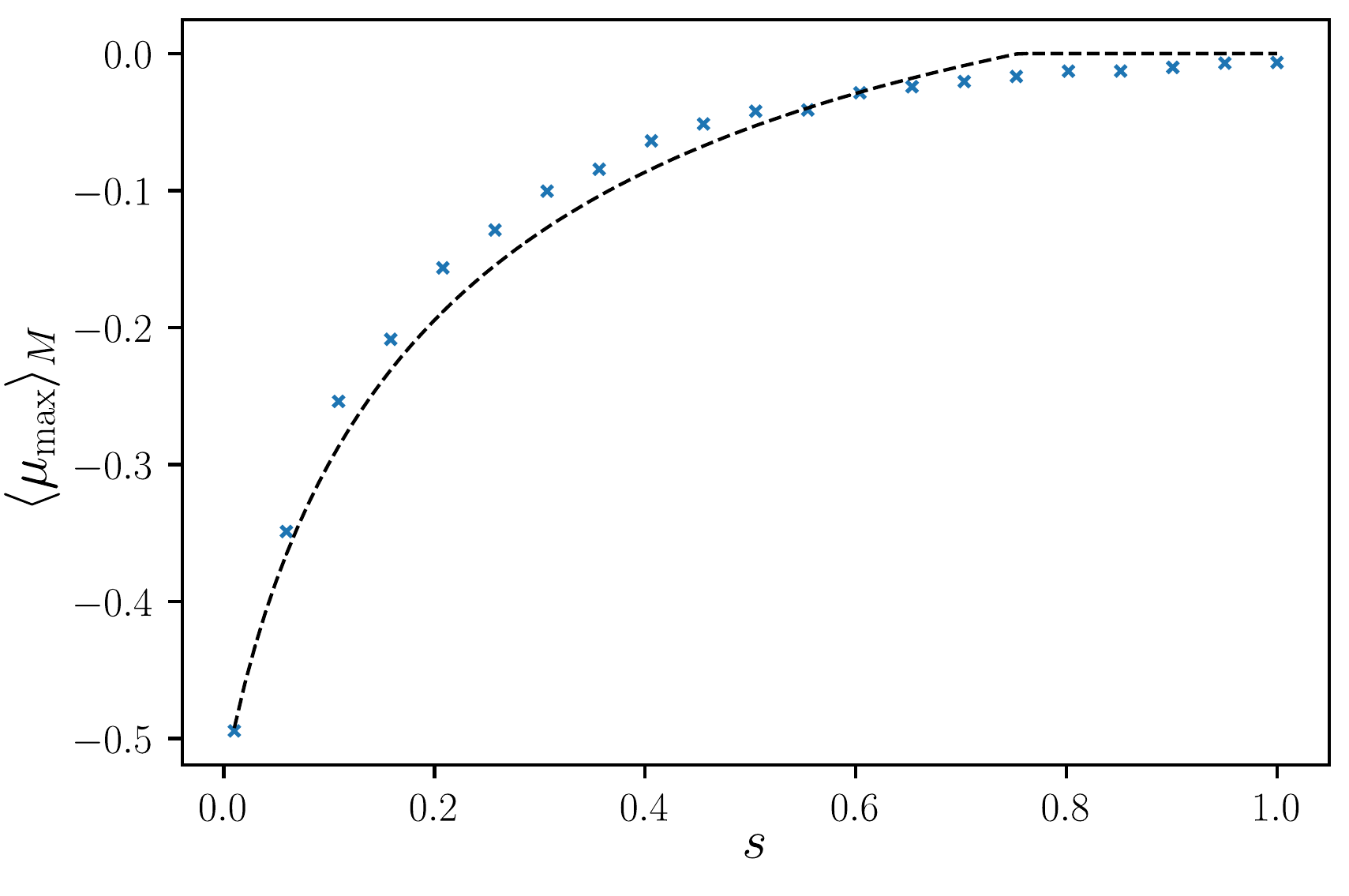}
	\caption{Expected maximum eigenvalue as a function of $s$. The remaining system parameters are $m = 1$, $f = 0.5$, $p = 10$ and $N = 100$. The blue crosses indicate the results of numerical diagonalisation averaged over $64$ trials. The dashed line is the upper bound of the EMA prediction. The critical value at which the EMA boundary hits zero for this value of $p$ is $s \approx 0.72$. } \label{fig:mumaxversuss}
\end{figure}

Although one can obtain a reasonable approximation of the bulk of the eigenvalue spectrum using the EMA, one cannot obtain the SDA in the same way as in Section \ref{sec:sdaer}. This is because the multi-dimensional integrals in Eq.~(\ref{sdasummary}) cannot be evaluated in a systematic way. However, the nature of these integrals tells us that no isolated peaks are predicted outside the bulk region of the spectrum, in contrast to the example in Section \ref{sec:sdaer}. Instead, there are smooth tails. 

Despite not being able to evaluate the SDA, we can still approximate the ELE using the EMA (see Fig.~\ref{fig:mumaxversuss}). The upper bound of the EMA works best as an approximation to the ELE when $N$ is low, but not so low as to invalidate the saddle-point approximation. If $N$ were to be increased from the value used in Fig.~\ref{fig:mumaxversuss}, the ELE would move further into the tail of the distribution and thus closer to zero, as is exemplified in Fig.~\ref{fig:mumaxversusn}. The upper bound of the EMA therefore acts as an approximate lower bound for the ELE.

We see in Fig.~\ref{fig:mumaxversuss} that increasing the variance of the coupling constants between voters (whilst holding the mean constant) can bring the largest eigenvalue arbitrarily close to zero. In fact, there is a value of $s$ where the boundary of the EMA hits $\mu= 0$. This critical value of $s$ increases as $p$ is increased, since greater connectivity reduces the ELE (see Fig. \ref{fig:mumaxversusp}). 

\section{Discussion and conclusion}\label{sec:discussion}
One point that this work was intended to highlight was that not all aspects of the dynamics in binary-state models can be captured by the statistics of macroscopic (population-averaged) order parameters. We have shown that microscopic order in the form of individual bias can persist even in regimes where there is no macroscopic order [as is exemplified in Fig. \ref{fig:individualbased}]. In more sociological terms, polarisation in the form of two sets of intransigent voters is possible and one can only detect this through the statistics of individuals.

After observing the possibility for individuals to have a long-lasting proclivity for one opinion over the other, we then went on to identify the factors that contributed to this effect. It was found that a low average connectivity $p$, a high degree of variation in interaction strength $s$ and a large population size $N$ encouraged persistent bias. That is, when individuals are influenced by a select few others (whom they may be influenced to imitate or contradict) and when a minority of this small group are more influential than the rest, individuals can form cliques of influence and biases can persist. A greater population size $N$ gives rise to a greater possibility for such cliques. 

It was the eigenvalue spectrum of the matrix $\hat M$ [defined in Eq.~(\ref{matdef})] that allowed us to quantify the persistence of individual bias. There are a range of analytical tools which one can use to evaluate the eigenvalue spectral density of a sparse random matrix. We chose to use the effective-medium and single-defect approximations here due to their ability to provide convenient closed-form expressions. One downside to this approach is that the results are not exact (although they can be very accurate) and the approximations are not necessarily well-controlled; one cannot precisely predict when they will fail and to what degree. 

One could instead have used the series expansion technique developed by Rodgers and Bray \cite{rodgersbray, brayrodgers}, the cavity approach of Rogers et al \cite{rogers2008cavity} or the stochastic population dynamics approach of K\"{u}hn \cite{kuhn2008spectra}. While the former method provides a more rigorous scheme of approximation and the latter two provide a near-perfect replication the spectrum, none of these approaches offer the closed-form expressions that we sought. It was such expressions that allowed the efficient evaluation of the leading eigenvalue as a function of the various system parameters. 

Using the EMA and SDA approximations to the eigenvalue spectrum of $\hat M$, we demonstrated that changing the value of $f$ (the fraction of antagonistic links) has the effect of changing only the location of one eigenvalue in the spectrum of $\hat M$. The fraction of antagonistic links thus has little effect on the microscopic statistics in many circumstances. With that being said, a non-zero value of $f$ does have an important effect on the macroscopic dynamics: the presence of antagonistic links means that a consensus cannot be reached. 

Microscopic order in the face of macroscopic disorder is reminiscent of the spin glass phase in magnetic materials. In fact, the spin-glass order parameter originally proposed by Edwards and Anderson \cite{edwards1975theory} (which was later refined \cite{sommers1982static}) can be seen to be non-zero when there is an eigenvalue of $\hat M$ which approaches zero [see Appendix \ref{appendix:correlations}]. Although we cannot identify precisely a critical point at which the leading eigenvalue tends to 0 in thermodynamic limit due to the nature of the approximations used here, we can at least say that the persistent bias in our model is akin to the behaviour of a spin-glass.

In summary, when the social network is sparsely connected (low $p$) and individuals are inclined to trust/oppose the opinions of a select few of their neighbours preferentially (high $s$), individuals are able to maintain a bias towards one opinion over the other. Such individual bias can persist even when the there is no macroscopic bias in the system and no consensus is reached. In cases where the eigenvalue spectrum of $\hat M$ extends to zero, the leading eigenvalue of the system will tend towards zero in the thermodynamic limit $N\to \infty$. This indicates that, in the thermodynamic limit, biases may persist for arbitrarily long times, even when an absorbing state is not reached.

An interesting avenue for future work would be to study this effect on different types of complex network (such as scale-free \cite{barabasi2003scale} or small-world networks \cite{watts1998collective, strogatz2001exploring}) or perhaps to make analytical headway with the more challenging model discussed in Appendix \ref{appendix:unweighted}. Work investigating the effect of quenched disorder on the macroscopic dynamics in the voter model in more detail is under way.

\acknowledgements
Partial financial support has been received from the Agencia Estatal de Investigaci\'on (AEI, MCI, Spain) and Fondo Europeo de Desarrollo Regional (FEDER, UE), under Project PACSS (RTI2018-093732-B-C21) and the Maria de Maeztu Program for units of Excellence in R\&D (MDM-2017-0711).

\begin{widetext}

\begin{appendix}

\section{Relationship between eigenvalues and long-term correlations}\label{appendix:correlations}
Calculating the single-site bias $m_i$ involves taking the ensemble average over all possible system trajectories for a given starting condition. Using a the more explicit notation $m_i(t | \{ s_j(0)\})$, we have
\begin{align}
m_i(t | \{ s_j(0)\}) = \sum_{s_i(t)} P(s_i(t) | \{ s_j(0)\}) [2 s_i(t) -1],
\end{align}
where $\{ s_j(0)\}$ is the starting configuration of the whole system. One may wish to use a quantity that reflects the degree to which single-site bias is preserved over time but that is independent of initial conditions. One can write an expression for the correlation of a single opinion at time $t$ with its starting value as follows
\begin{align}
C_i(t,0) &\equiv \langle \langle [2 s_i(t) - 1][2 s_i(0) -1]\rangle \rangle - \langle \langle 2 s_i(t) - 1 \rangle\rangle \langle \langle 2 s_i(0) -1 \rangle\rangle \nonumber \\
\end{align}
where the double angular brackets represent an average over the starting configuration and all possible resulting trajectories. Assuming a symmetrical initial distribution of biases, we see that the terms $\langle\langle 2 s_i(t) - 1 \rangle\rangle$ and $\langle\langle 2 s_i(0) - 1 \rangle\rangle$ vanish and one obtains
\begin{align}
C_i(t,0) &= \sum_{s_i(t), \{ s_j(0)\} } P(s_i(t) | \{ s_j(0)\})\, P(\{ s_j(0)\})\, [2 s_i(t) - 1][2 s_i(0) -1]  \nonumber \\
&= \sum_{\{ s_j(0)\}} m_i(t | \{ s_j(0)\})\, P(\{ s_j(0)\}) \, [2 s_i(0) -1]   . 
\end{align}
 We can replace $m_i(t, \{ s_j(0)\})$ using the eigenbasis decomposition in Eq.~(\ref{eigenbasisdecomp}) to write
\begin{align}
C_i(t,0) &=  \sum_{\mu} \sum_{\{ s_j(0)\} }  P(\{ s_j(0)\}) \, [2 s_i(0) -1] \, c_{(\mu)}(\{ s_j(0)\}) v^{(\mu)}_i e^{-\mu t}   .
\end{align}
Since $C_i(0,0)>0$, we see that when there is an eigenvalue $\mu$ which approaches 0, biases can have a non-vanishing correlation with their starting values for large $t$. The original order parameter proposed by Edwards and Anderson for the study of spin glasses was $\lim_{t \to \infty}C_i (t,0)$ \cite{edwards1975theory}. 

\section{Model with node-oriented update procedure}\label{appendix:unweighted}
In the model presented in Section \ref{sec:modeldef}, links are chosen at a rate proportional to their weight and, if the link is active, one of the voters changes its opinion. An equivalent node-centred formulation of these dynamics would be to select a node proportional to it's degree and then choose a neighbour to copy with probability proportional to the weight of the link. 

In most formulations of the voter model on complex networks, nodes are selected to change at a rate independent of their degree. To check that the qualitative predictions of our model carry over to this case, we repeat the numerical simulations used to produce Figs. \ref{fig:mumaxversusn}, \ref{fig:mumaxversusp} and \ref{fig:mumaxversuss} using a matrix of the form 
\begin{align}
M_{ij} = \frac{p}{p_i}(2 A_{ij} -1)J_{ij}  - \delta_{ij} \frac{p}{p_i}\sum_k J_{ik} ,\label{altrates}
\end{align}
where $p_i$ is the degree of node $i$. We see that although the ELE is quantitatively different (as should be expected), similar dependencies of the ELE on the system parameters $N$, $p$ and $s$ carry over to this case. This suggests that our general conclusions regarding the influence of system size, link paucity and link weight variation carry over to this case. 

\begin{figure}[H]
	\centering 
	\includegraphics[scale = 0.36]{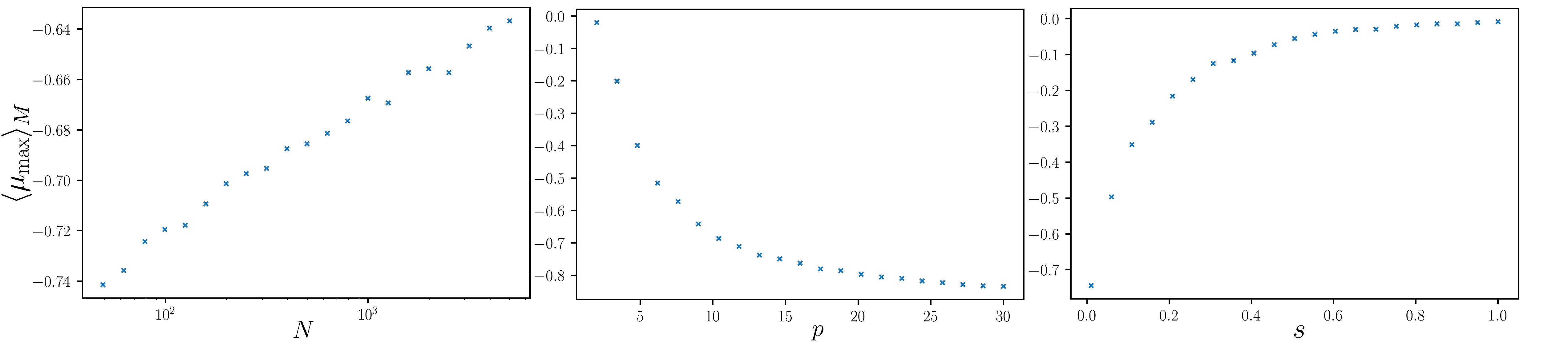}
	\caption{Reproduction of Figs. \ref{fig:mumaxversusn}, \ref{fig:mumaxversusp} and \ref{fig:mumaxversuss} using matrices of the form Eq.~(\ref{altrates}) instead of Eq.~(\ref{matdef}), with all other model parameters being equal. } \label{fig:altrates}
\end{figure}

\section{Derivation of the expressions for the EMA and SDA} \label{appendix:summary}
\subsection{General formulation and setting up the saddle-point problem}
This appendix summarises how one arrives at the general expressions for the effective-medium approximation (EMA) and the single-defect approximation (SDA) for the eigenvalue density of the matrix $\hat M$, Eqs.~(\ref{emasummary}) and (\ref{sdasummary}) respectively. Following standard methods \cite{semerjian, edwardsjones, sommers, barongalla}, one can find the density of eigenvalues $\rho(\mu)$ via the resolvent. We have
\begin{align}
\rho(\mu) = \frac{1}{N\pi} \mathrm{Im} \,\mathrm{Tr} \left[\hat M - (\mu + i \epsilon) \id \right]^{-1} = \frac{2}{N \pi} \mathrm{Im} \frac{\partial \ln Z}{\partial \mu} ,
\end{align}
where $\epsilon$ is a small real number and we have defined the partition function $Z(\mu)$, which can be written as a Gaussian integral 
\begin{align}
&Z(\mu) = \int_{-\infty}^{\infty} \left[ \prod_i d \phi_i \right] \exp\left[ \frac{i}{2} \left( \mu\sum_i \phi_i^2 - \sum_{ij} M_{ij} \phi_i \phi_j \right)\right].
\end{align}
We are interested in the disorder-averaged eigenvalue density, as opposed to the eigenvalue density of any one particular randomly drawn matrix. In order to take the average over the ensemble of random matrices (indicated by $[\cdot]_M$), we employ the replica trick \cite{mezard1987} $[\ln Z]_{M} = \lim_{n \to 0} ([Z^n]_{M}- 1)/n$, where
\begin{align}
Z(\mu)^n = \int_{-\infty}^{\infty} \left[ \prod_{i, \alpha} d \phi^\alpha_i \right] \exp\left[ \frac{i}{2} \left( \mu \sum_{i,\alpha} (\phi^\alpha_i)^2 - \sum_{ij \alpha} M_{ij} \phi^\alpha_i \phi^\alpha_j \right)\right] , \label{replicated}
\end{align}
and where $\alpha$ indexes the $n$ replicas. We now introduce the quantity $c(\boldphi)$, which is related to the fraction of sites with a field $\boldphi_i$ equal to a reference value $\boldphi$, where the bold script indicates a vector in replica space
\begin{align}
c(\boldphi) = \frac{1}{N} \sum_i \delta\left(\boldphi - \boldphi_i\right). \label{boldphidef}
\end{align}
One therefore obtains for the partition function \cite{semerjian, nagao2007spectral}
\begin{align}
[Z(\lambda)^n]_M =& \int \mathcal{D} c(\boldphi) \exp(-N S_{\mathrm{eff}}) , \nonumber \\
S_{\mathrm{eff}}\left[ c(\boldphi)\right] =& \int d\boldphi c(\boldphi) \ln c(\boldphi) \nonumber \\
&- \frac{i\mu}{2} \int d\boldphi c(\boldphi) \boldphi^2 + H_{\mathrm{eff}}\left[ c(\boldphi)\right] ,\label{effectiveaction}
\end{align}
where the integration is taken only over normalised $c(\boldphi)$ such that $\int \mathcal{D} c(\boldphi) = 1$ and we have defined the effective Hamiltonian
\begin{align}
\exp\left[ -N H_{\mathrm{eff}}\right] = \left[\exp\left(-\frac{i}{2} \sum_{ij} M_{ij} \boldphi_i \cdot \boldphi_j \right) \right]_{M} . \label{effham}
\end{align}
We note that the entropic contribution to $S_{\mathrm{eff}}$ in Eq.~(\ref{effectiveaction}) arises from the possible combinations of sites occupying the state $\boldphi$ \cite{monasson} (see Refs. \cite{nagao2007spectral, nagao2008spectral} for a more thorough account of this term). The eigenvalue density is recovered from Eq.~(\ref{effectiveaction}) via
\begin{align}
\rho(\mu) &= \lim_{n\to 0} \frac{2}{n N \pi} \mathrm{Im} \frac{\partial}{\partial \mu} \ln \left[ Z^n\right]  \nonumber \\
&= \lim_{n \to 0} \frac{1}{\pi n} \mathrm{Re} \int d \boldphi c(\boldphi) \boldphi^2 . \label{recovereigenvaluedensity}
\end{align}
Our strategy is now to evaluate the effective action $S_{\mathrm{eff}}$ in the limit $N~\to~\infty$, using the saddle-point approximation, with $p$ held finite. The saddle-point satisfies the condition $\frac{\delta S_{\mathrm{eff}}}{\delta c(\boldphi)} = 0$, or more explicitly
\begin{align}
c(\boldphi) = \mathcal{N} \exp\left[ \frac{i}{2} \mu \boldphi^2 - \frac{\delta H_{\mathrm{eff}}}{\delta c(\boldphi)}\right], \label{saddlepoint}
\end{align}
where $\mathcal{N}$ is a normalisation constant. In general, this cannot be solved exactly and one has to resort to a further approximation scheme. Two related approximations (the EMA and SDA), each of which has its own advantages and limitations, are described below. We note that, up until this point, our consideration has been entirely general and we have not used a specific form of the matrix $\hat M$.

\subsection{The effective-medium approximation (EMA)}\label{appendix:emasummary}
Because each lattice site $i$ is statistically equivalent and the total number of sites is large, we can make the approximation that the number of sites taking a particular field value $\boldphi$ is approximately Gaussian \cite{semerjian}. This is expected to be a valid approximation provided $p$ is sufficiently large. We therefore employ the ansatz
\begin{align}
c^{\mathrm{EMA}}(\boldphi) = (2 \pi i \sigma(\mu))^{-n/2} \exp\left( - \frac{\boldphi^2}{2 i \sigma(\mu)}\right). \label{emaansatz}
\end{align}
We note that in this ansatz the replicas are independent and symmetric with respect to one another. The eigenvalue density is thus given by [using Eq.~(\ref{recovereigenvaluedensity})]
\begin{align}
\rho^{\mathrm{EMA}}(\mu) = - \frac{1}{\pi} \lim_{n \to 0} \mathrm{Im} \, \sigma(\mu + i\epsilon) , \label{spectrumfromsigma}
\end{align}
where $\epsilon$ is a small real quantity introduced to avoid any singularities \cite{sommers}. Substituting Eq.~(\ref{emaansatz}) into Eq.~(\ref{effectiveaction}), we obtain a far simpler saddle-point equation $d S_{\mathrm{eff}}/d \sigma(\mu) =0$, or more precisely
\begin{align}
\frac{\mu}{2}-\frac{1}{2 \sigma} + \frac{d H_{\mathrm{eff}}}{d \sigma} = 0.
\end{align}
\subsubsection{General $\pi(J)$}
Taking the general case where the random matrix $M_{ij}$ is as defined in Eq.~(\ref{matdef}) and Eq.~(\ref{sparsedef}), one obtains for the effective Hamiltonian defined in Eq.~(\ref{effham}) 
\begin{align}
-H_{\mathrm{eff}} = -\frac{p}{2} + \frac{p}{2}\int dJ \pi(J) \int d \boldphi d \boldpsi \, c(\boldphi) c(\boldpsi) \left\{f \exp\left[ \frac{iJ}{2} (\boldphi+ \boldpsi)^2\right] +(1-f) \exp\left[ \frac{iJ}{2} (\boldphi- \boldpsi)^2\right] \right\}, \label{hamiltonianstrict}
\end{align}
where we have used the definition of $c(\boldphi)$ in Eq.~(\ref{boldphidef}) and we have used the result $1 + a/N \approx \exp(a/N)$ for large $N$ to re-exponentiate. Using the EMA ansatz in Eq.~(\ref{emaansatz}), the integration over the fields $\boldphi$ and $\boldpsi$ can be performed explicitly to yield
\begin{align}
\int d \boldphi d \boldpsi \, c(\boldphi) c(\boldpsi) \exp\left[iJ \frac{ (\boldphi - \boldpsi)^2}{2} \right] &= \left\{\int d \phi^\alpha d\psi^\alpha \frac{1}{2 \pi i \sigma}\exp\left[ -\frac{(\phi^\alpha)^2 + (\psi^\alpha)^2}{2 i \sigma}\right]\exp\left[ i J \frac{\left(\phi^\alpha-\psi^\alpha\right)^2}{2 }\right] \right\}^n \nonumber \\
&=  (-i)^n\left[ 1 + 2\sigma J\right]^{-\frac{n}{2}},
\end{align}
and 
\begin{align}
\int d \boldphi d \boldpsi \, c(\boldphi) c(\boldpsi) \exp\left[i J\frac{ (\boldphi + \boldpsi)^2}{2} \right] &= \left\{\int d \phi^\alpha d\psi^\alpha \frac{1}{2 \pi i \sigma}\exp\left[ -\frac{(\phi^\alpha)^2 + (\psi^\alpha)^2}{2 i \sigma}\right]\exp\left[ i J \frac{\left(\phi^\alpha+\psi^\alpha\right)^2}{2 }\right] \right\}^n \nonumber \\
&= (-i)^n \left[ 1 + 2\sigma J\right]^{-\frac{n}{2}} .
\end{align}
Importantly we note that the dependence on $f$ vanishes. That is, the bulk of the eigenvalue spectrum (in the EMA approximation) is independent of the proportion of antagonistic connections. From Eq.~(\ref{effectiveaction}), we have
\begin{align}
S_{\mathrm{eff}}\left[ \sigma\right] =& -\frac{n}{2}\ln(2 \pi i \sigma) - \frac{n}{2} + \frac{n \mu}{2} \sigma  + \frac{p}{2} - \frac{p}{2} \int dJ \, \pi(J) \left[ 1 + 2\sigma J\right]^{-\frac{n}{2}} .
\end{align}
We are now in a position to solve the saddle-point equation $d S_{\mathrm{eff}}/d \sigma =0$ and take the limit $n\to 0$. This yields the function $\sigma(\mu)$ that can be related to the eigenvalue density via Eq.~(\ref{spectrumfromsigma}). Differentiating with respect to $\sigma$ and taking the limit $n\to 0$ one obtains
\begin{align}
\lim_{n\to 0 }\frac{1}{n}\frac{\delta S_{\mathrm{eff}}}{\delta \sigma} =\frac{\mu}{2}-\frac{1}{2 \sigma} + \frac{p}{2} \int dJ \, \pi(J) \frac{J}{ 1 + 2\sigma J} . \label{diffheffer}
\end{align}
Setting the right-hand side equal to zero and rearranging, one arrives at the second of Eqs.~(\ref{emasummary}) in the main text. We recognise the integral in Eq.~(\ref{emasummary}) as being related to the Stieltjes transform of the probability distribution $\pi(J)$. If the Stieltjes transform of $\pi(J)$ is known, then Eq.~(\ref{emasummary}) can be inverted for $\sigma(\mu)$, which can then be used to yield the eigenvalue density via Eq.~(\ref{spectrumfromsigma}).

Although the EMA replicates the main `bulk' of the eigenvalue distribution fairly well, it does a poor job of recovering the outlier eigenvalues which come about due to localisation effects \cite{birolisda, Bray_1982}. For this reason, it is beneficial to go one level of approximation further by substituting the EMA back into the original saddle-point solution of Eqs.~(\ref{effectiveaction}) [see Appendix \ref{appendix:sdasummary} on the SDA].

\subsubsection{Special case: $\pi(J) = \delta(J-1/p)$}
Substituting $\pi(J) = \delta(J - 1/p)$ in Eq.~(\ref{emasummary}) and solving for $\sigma(\mu)$, one obtains
\begin{align}
\sigma(\mu) &= \frac{2 - (\mu + 1) p - \sqrt{8 \mu p + (p(\mu+1) -2)^2}}{4\mu}, \label{emastrict}
\end{align}
Substituting Eq.~(\ref{emastrict}) into Eq.~(\ref{spectrumfromsigma}), one ultimately arrives at Eq.~(\ref{emalaplacian}). One can show that the expression in Eq.~(\ref{emalaplacian}) is normalised by making the change of variables $x = \sqrt{\frac{p}{8}}(1+\mu) + \sqrt{\frac{1}{2p}}$ and by using the result $\int dx \sqrt{1-x^2}/(a-x) = -\sqrt{1-x^2} + a \arcsin(x) - \sqrt{1-a^2} \arctan\left[(ax-1)/(\sqrt{1-a^2}\sqrt{1-x^2})\right]$.
 
\pagebreak

\subsection{The single-defect approximation}\label{appendix:sdasummary}
The single-defect approximation involves taking the Gaussian ansatz for the EMA, and substituting this back into the original saddle-point condition in order to obtain a more accurate expression for the eigenvalue density. In principle, this resubstitution procedure can be iterated many times. Although this produces a more accurate approximation of the eigenvalue spectrum, the expressions one obtains can become more unwieldy and therefore less useful. 

More specifically, returning to the saddle-point condition Eq.~(\ref{saddlepoint}) and expanding this as a series, one obtains
\begin{align}
c(\boldphi) = \mathcal{N} e^{\frac{i}{2} \mu \boldphi^2} \sum_{k=0}^\infty \left[- \frac{\delta H_{\mathrm{eff}}}{\delta c(\boldphi)} \right]^k . \label{sdaapproxgeneral}
\end{align}
One now inserts the EMA solution into the right-hand side of this equation and uses Eq.~(\ref{recovereigenvaluedensity}) to obtain the single-defect approximation to the eigenvalue density \cite{birolisda, semerjian}. 

In order to evaluate Eq.~(\ref{sdaapproxgeneral}), we must first find $\frac{H_{\mathrm{eff}}}{\delta c(\boldphi)}$. Differentiating Eq.~(\ref{hamiltonianstrict}) and exploiting the independence from $f$, one obtains
\begin{align}
-\frac{H_{\mathrm{eff}}}{\delta c(\boldphi)} &=  p\int dJ\, \pi(J)  \int  d \boldpsi \, c(\boldpsi) \left\{\exp\left[ \frac{i J}{2} (\boldphi+ \boldpsi)^2\right]  \right\} \nonumber \\
&= p\int dJ\, \pi(J)  \prod_\alpha \left\{\int d \psi^\alpha \frac{1}{\sqrt{2 \pi i \sigma}}\exp\left[ -\frac{ (\psi^\alpha)^2}{2 i \sigma}\right]\exp\left[ iJ \frac{\left(\phi^\alpha+\psi^\alpha\right)^2}{2 }\right] \right\} \nonumber \\
&= \int dJ\, \pi(J) \left[1 + J\sigma \right]^{-n/2}\exp\left[ \frac{i  \boldphi^2}{2 (1/J+\sigma)}\right] ,
\end{align}
where we have inserted the Gaussian ansatz Eq.~(\ref{emaansatz}) to evaluate the integrals over $\boldpsi$. This can now be substituted into Eq.~(\ref{sdaapproxgeneral}) to give
\begin{align}
c(\boldphi) = \mathcal{\eta} e^{\frac{i}{2} \mu \boldphi^2} \sum_{k=0}^\infty \frac{e^{-p} p^k}{k!} \left\{\int dJ \, \pi(J)\left[1 + J\sigma \right]^{-n/2} \exp\left[ \frac{i  \boldphi^2}{2 (1/J+\sigma)}\right] \right\}^k , \label{sdageneralc}
\end{align}
where we see that the normalisation constant $\eta = \mathcal{N} e^p$ tends to unity when $n \to 0$. Finally, using Eq.~(\ref{recovereigenvaluedensity}), one obtains the final expression for the SDA. One notes that the integral over $J$ and the limit $n \to 0$ are interchangeable (this can be shown by integrating by parts). Upon substitution of Eq.~(\ref{sdageneralc}) into Eq.~(\ref{recovereigenvaluedensity}), the general expression for the SDA of the eigenvalue density is then given by Eq.~(\ref{sdasummary}).

\end{appendix}

\end{widetext}

\end{document}